\documentclass[preprints,article,accept,pdftex,moreauthors]{Definitions/mdpi} 

\firstpage{1} 
\pubvolume{1}
\issuenum{1}
\articlenumber{0}
\pubyear{2023}
\copyrightyear{2023}
\datereceived{ } 
\daterevised{ } 
\dateaccepted{ } 
\datepublished{ } 
\hreflink{https://doi.org/} 

\graphicspath{{./images/}}
\usepackage{caption}
\usepackage{subcaption}
\DeclareCaptionLabelFormat{subcaptionlabel}{\normalfont(\textbf{#2}\normalfont)\\}
\Title{Size Effect on the Thermal Conductivity of a Type-I Clathrate}

\TitleCitation{Size Effect on the Thermal Conductivity of a Type-I Clathrate}

\Author{Monika Lu\v{z}nik $^{1,}$*, Günther Lientschnig $^{1}$, Mathieu Taupin $^{1}$, Andreas Steiger-Thirsfeld $^{2}$, Andrey Prokofiev ${^1}$\linebreak  and Silke Paschen $^{1,}$*\orcidA{}}

\AuthorNames{Monika Lu\v{z}nik, Günther Lientschnig, Mathieu Taupin, Andreas Steiger-Thirsfeld, Andrey Prokofiev and Silke Paschen}

\AuthorCitation{Lu\v{z}nik, M.; Lientschnig, G.; Taupin, M.; Steiger-Thirsfeld, A.; Prokofiev, A.; Paschen, S.}

\address{%
$^{1}$ \quad Institute of Solid State Physics, TU Wien, Wiedner Hauptstraße 8-10, 1040 Wien, Austria\\
$^{2}$ \quad USTEM, TU Wien, Wiedner Hauptstraße 8-10, 1040 Wien, Austria}

\corres{Correspondence: budnowski@ifp.tuwien.ac.at (M.L.); paschen@ifp.tuwien.ac.at (S.P.)}

\abstract{Clathrates are a materials class with an extremely low phonon thermal conductivity, which is a key ingredient for a high thermoelectric conversion efficiency. Here, we present a study on the type-I clathrate La$_{1.2}$Ba$_{6.8}$Au$_{5.8}$Si$_{38.8}\square_{1.4}$ directed at lowering the phonon thermal conductivity even further by forming mesoscopic wires out of it. Our hypothesis is that the interaction of the low-energy rattling modes of the guest atoms (La and Ba) with the acoustic modes, which originate mainly from the type-I clathrate framework (formed by Au and Si atoms, with some vacancies $\square$), cuts off their dispersion and thereby tilts the balance of phonons relevant for thermal transport to long-wavelength ones. Thus, size effects are expected to set in at relatively long length scales. The structuring was carried out using a top-down approach, where the wires, ranging from 1260~nm to 630~nm in diameter, were cut from a piece of single crystal using a focused ion beam technique. Measurements of the thermal conductivity were performed with a self-heating $3\omega$ technique down to 80~K. Indeed, they reveal a reduction of the room-temperature phonon thermal conductivity by a sizable fraction of $\sim$40\,\% for our thinnest wire, thereby confirming our hypothesis.
}


\begin{document}

\section{Introduction}

In view of global warming and the recent energy crisis, sustainable alternatives to fossil fuels are more important than ever \cite{IEA2022, Dong2022}. Whereas at present thermoelectric generators (TEGs) play only a minor role in the wide group of renewable energy sources, their ability to turn a temperature difference into electricity makes them uniquely suited to be used in combination with other generators to recover energy that is otherwise lost as waste heat~\cite{Pourkiaei2019}. To boost the energy output, however, it is necessary to maximize their thermoelectric conversion efficiency. This in turn requires the used materials to have a high thermoelectric figure of merit $ZT$.
\vspace{-6pt}

\begin{figure}[H]
	\begin{subfigure}[t]{0.45\textwidth}
        \subcaption{}
        \includegraphics[height=5cm]{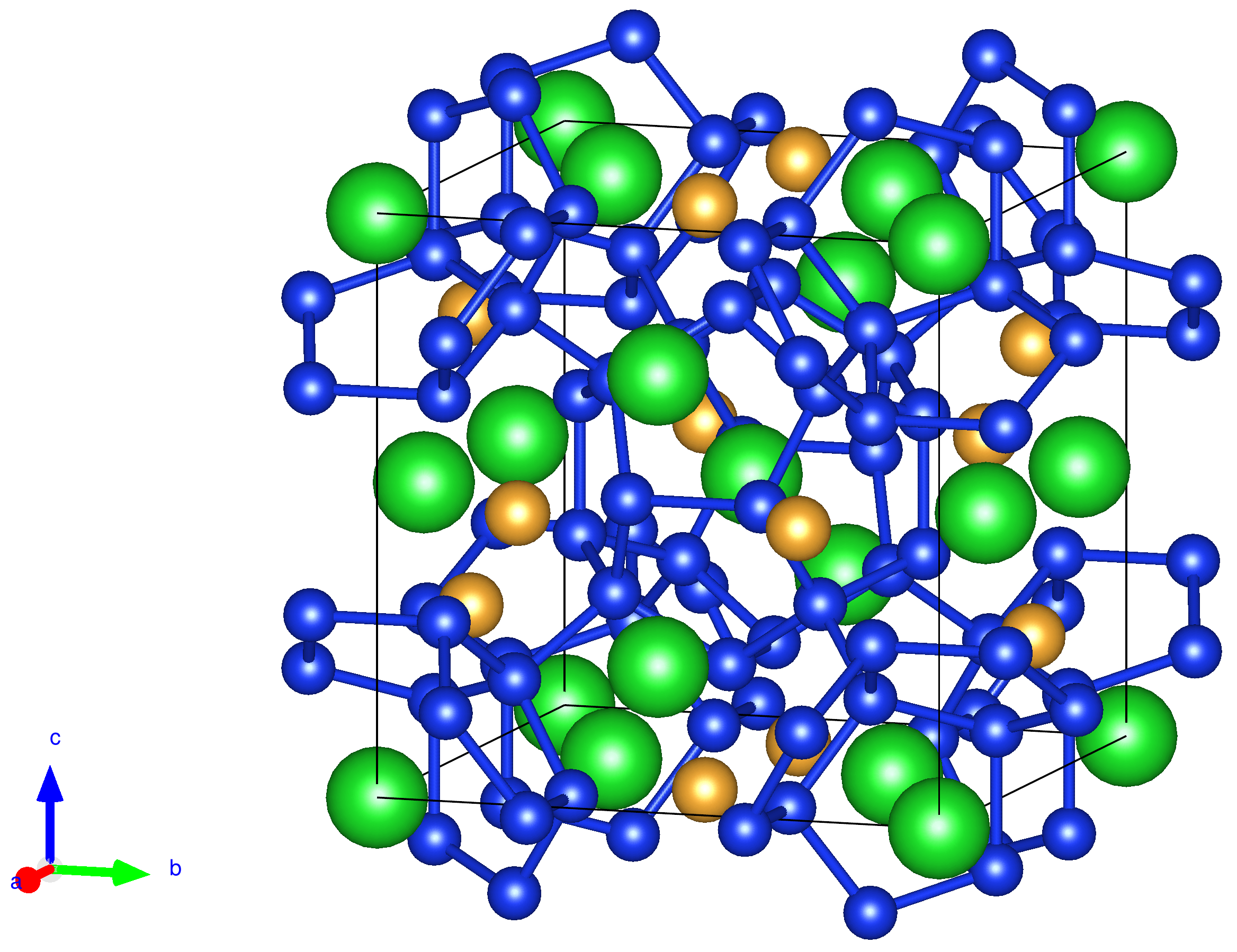}
		\label{fig:crystal structure}
	\end{subfigure} \quad
	\begin{subfigure}[t]{0.35\textwidth}
        \subcaption{}
		 \includegraphics[height=5cm]{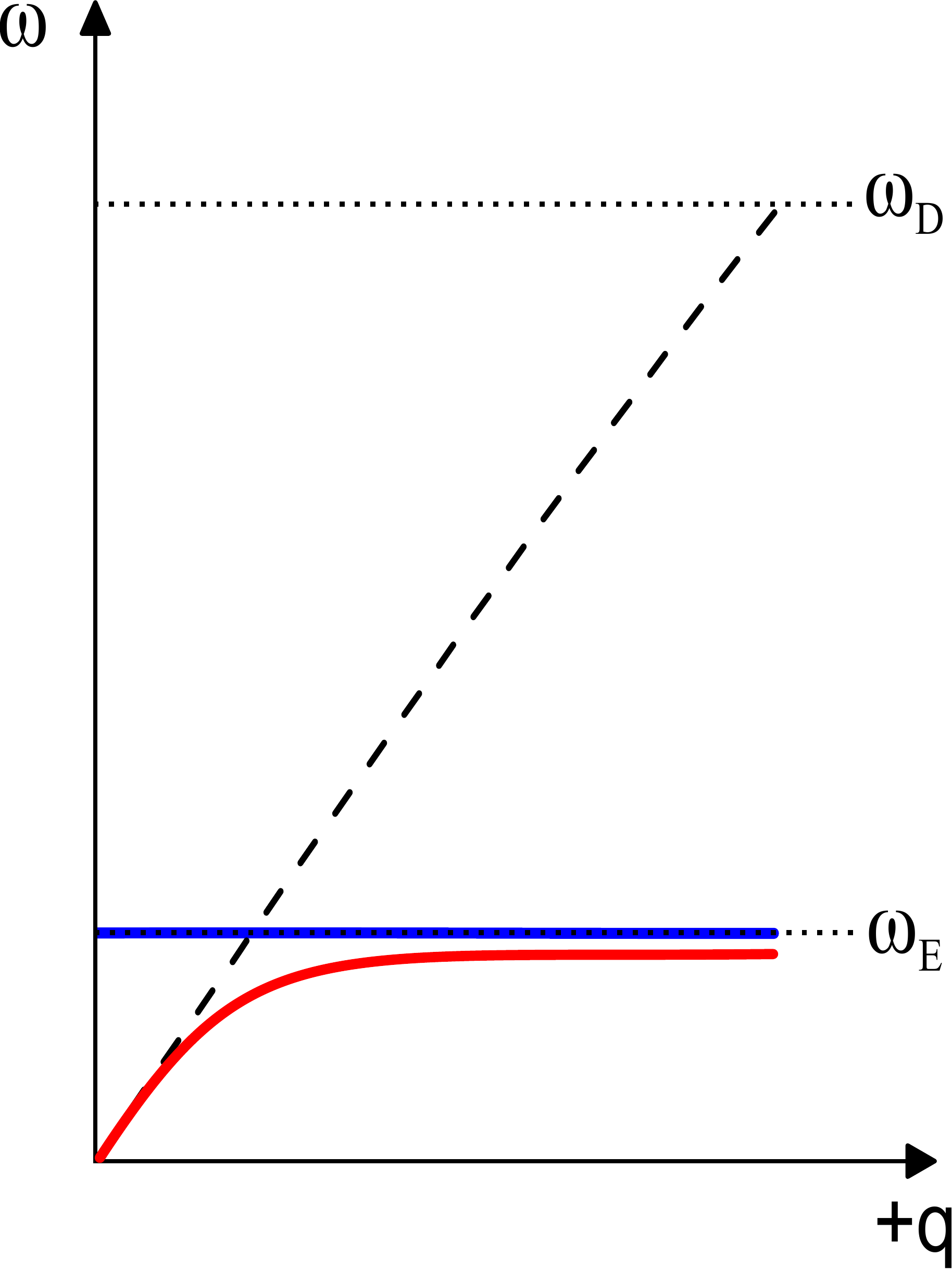}
		\label{fig:dispersion_sketch}
	\end{subfigure}
	\caption{(\textbf{a}) Crystal structure \cite{Momma2011} of a type-I clathrate, with the guest atoms (green) sitting inside cages formed by essentially covalently bonded host atoms (blue and orange). In the case of La$_{1.2}$Ba$_{6.8}$Au$_{5.8}$Si$_{38.8}\square_{1.4}$ (La-BAS), the green spheres denote Ba and La, the blue ones Si and the orange ones Au. The vacancies $\square$ are distributed between Si and Au; (\textbf{b}) sketch of the phonon dispersion of type-I clathrates, as proposed in \cite{Ikeda2019}. An acoustic phonon mode (the dashed line shows the dispersion assumed in the Debye model) hybridizes with a flat rattling mode (blue), resulting in a severe flattening of the former at large wave vectors $q$ (red). This shifts the characteristic energy scale from the Debye energy $\hbar\omega_{\rm D}$ to the Einstein energy $\hbar\omega_{\rm E}$ of the lowest-lying rattling mode.}
	\label{fig1}
\end{figure}

A material's (dimensionless) thermoelectric figure of merit is given by $ZT=\dfrac{S^2\sigma}{\kappa}T$, with the absolute temperature $T$, the Seebeck coefficient $S$, the electrical conductivity $\sigma$ and the thermal conductivity $\kappa$. $S$, $\sigma$ and $\kappa$ are interrelated via their dependencies on the charge carrier concentration, which are such that maximizing $ZT$ amounts to a compromise~\cite{Snyder2008}. However, the fact that the thermal conductivity is determined by both an electronic and a phonon contribution opens up the possibility to reduce $\kappa$ without significantly degrading the electronic properties. This is the basis for the so-called phonon glass--electron crystal concept, which has triggered the research on several materials classes.
Clathrates, for instance, form cage-like structures that encapsulate heavy guest atoms (Figure~\ref{fig1}a). The rattling of these guest atoms disrupts the heat-carrying phonons, thus leading to a remarkably low phonon thermal conductivity, while leaving the electrical conductivity essentially unaffected \cite{Nol01.1,Pas01.2,Pas03.2,Ben04.1,Tak14.1}.

Using inelastic neutron and X-ray scattering, an avoided crossing between the acoustic phonon modes and the flat rattling modes was identified \cite{Christensen2008,Euc12.1,Pai14.1}, as illustrated in \mbox{Figure~\ref{fig1}b}. The strong interaction of these modes, together with the associated thermodynamic signatures, the particular temperature dependence of the thermal conductivity and the observed universal scaling of the room-temperature phonon thermal conductivity with the product of sound velocity and Einstein temperature of the lowest-lying rattling mode, were taken as evidence for a phonon Kondo effect as the underlying mechanism \cite{Ikeda2019}.
The suppression of high-frequency acoustic phonons not only causes remarkably low phonon thermal conductivities, but also means that low-frequency (long-wavelength) phonons have a greater share in the heat transport. In fact, calculations suggest that in certain clathrates more than 90\% of the heat is carried by phonons with a wavelength longer than 10~nm \cite{Kaur2019}.
Therefore, even mesoscopic structures (about 100~nm to a few $\upmu$m) made from those clathrates are expected to lead to a sizable reduction of the thermal conductivity.

We set out to test this conjecture by measuring individual type-I clathrate mesowires, fabricated from a clathrate single crystal using a focused ion beam (FIB). We study the electrical resistivity and thermal conductivity at temperatures between 300~K and 80~K, where the lattice thermal conductivity of bulk type-I clathrates was shown to be dominated by intrinsic phonon--phonon Umklapp scattering; notably, for two well-controlled sample series---one with a varying vacancy content and one with a varying charge carrier concentration---the curves collapsed above about 50~K, demonstrating that scattering of phonons from vacancies and electrons does not play a significant role at higher temperatures \cite{Ikeda2019}. To avoid uncertainties in the thermal conductivity measurement, which are common in this temperature range, we implement a $3\omega$ technique in which a suspended mesowire acts as both the heater and the thermometer \cite{Lu2001}. Our results confirm that, even with relatively thick wires (diameters $< 600$~nm), a sizable reduction of the phonon thermal conductivity is achieved.

A number of previous studies explored size effects on the thermoelectric properties of clathrates in polycrystalline bulk samples, using melt spinning \cite{Yan2009, Tomes2016, Christian2016}, ball milling and (high-pressure torsion) hot pressing \cite{Christian2016, Zolriasatein2015, Yan2013}, high-pressure and high-temperature (HPHT) sintering \cite{Sun2018, Sun2020}, as well as approaches that introduced a secondary phase to stabilize small grains \cite{Yan2020}. Whereas some of the results were promising, firm conclusions on the relation between the grain size and the lattice thermal conductivity remained elusive. This calls for a proof-of-concept study on single wires, as performed in the present work.

\section{Materials and Methods}

\subsection{Samples}
For our study, we chose La$_{1.2}$Ba$_{6.8}$Au$_{5.8}$Si$_{38.8}\square_{1.4}$ (La-BAS) single crystals, which were grown in-house using a floating zone technique. They were studied using scanning electron microscopy (SEM) with energy-dispersive X-ray spectroscopy (EDX) and wavelength-dispersive X-ray spectroscopy (WDX) analyses, transmission electron microscopy (TEM) as well as powder X-ray diffraction (XRD). This thorough analysis confirmed the phase purity of the grown crystals down to the nanometer scale \cite{Prokofiev2013}. As a result of the decreased charge carrier concentration, the substitution with La enhances the thermopower of the Ba$_{8}$Au$_{x}$Si$_{46-x}$ series substantially \cite{Prokofiev2013}. However, while the phonon thermal conductivity $\kappa_{\rm ph} \sim 1.8$\,Wm$^{-1}$K$^{-1}$ at 300~K is within the range of the Ba$_{8}$Au$_{x}$Si$_{46-x}$ series \cite{Candolfi2012}, it is rather high compared to other clathrates, leaving room for improvement. We thus found this well-characterized compound to be ideally suited for our mesostructuring study.

\begin{figure}[H]
	\begin{subfigure}[t]{0.47\textwidth}
		\subcaption{}
		\includegraphics[width=0.85\linewidth]{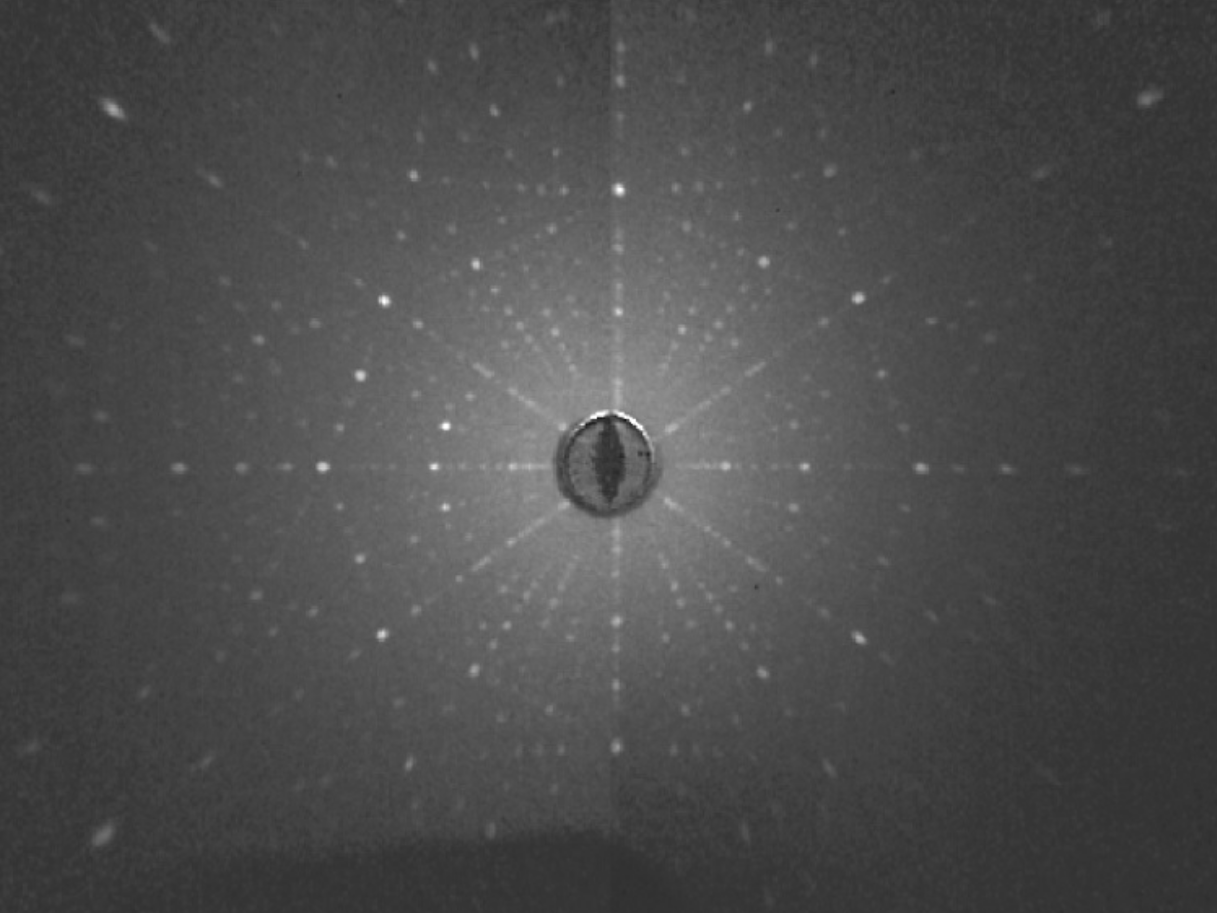}
		\label{fig:laue}
	\end{subfigure}
	\begin{subfigure}[t]{0.47\textwidth}
        \subcaption{}
		\includegraphics[width=0.85\linewidth]{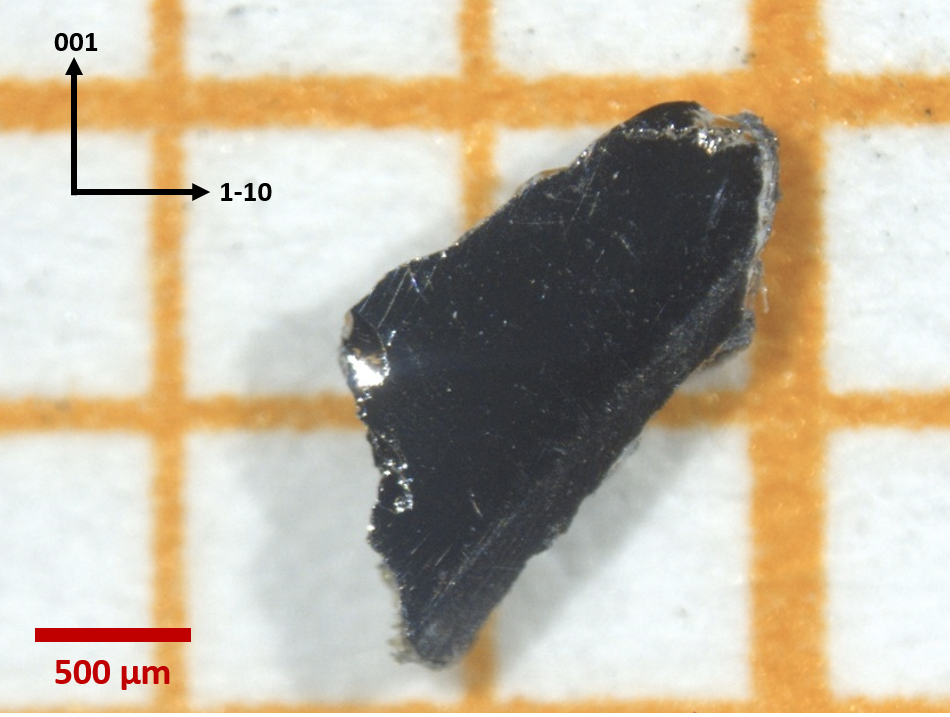}
		\label{fig:bulk}
	\end{subfigure}\\
	\caption{\textls[-8]{(\textbf{a}) Laue X-ray diffraction pattern of the used La-BAS single crystal; (\textbf{b}) oriented single~crystal}.}
	\label{fig2}
\end{figure}

A piece of crystal was investigated using Laue X-ray diffraction. We confirmed that it was indeed a single crystal (Figure \ref{fig2}a) and oriented it (Figure \ref{fig2}b). A crystal oriented along the 1--10 direction was then used to fabricate several mesowires with diameters of approximately 630~nm, 920~nm and 1260~nm.

\subsection{Nanofabrication Process}
\begin{figure}[H]
	\begin{subfigure}[t]{0.47\textwidth}
		\subcaption{}
		\includegraphics[width=0.85\linewidth]{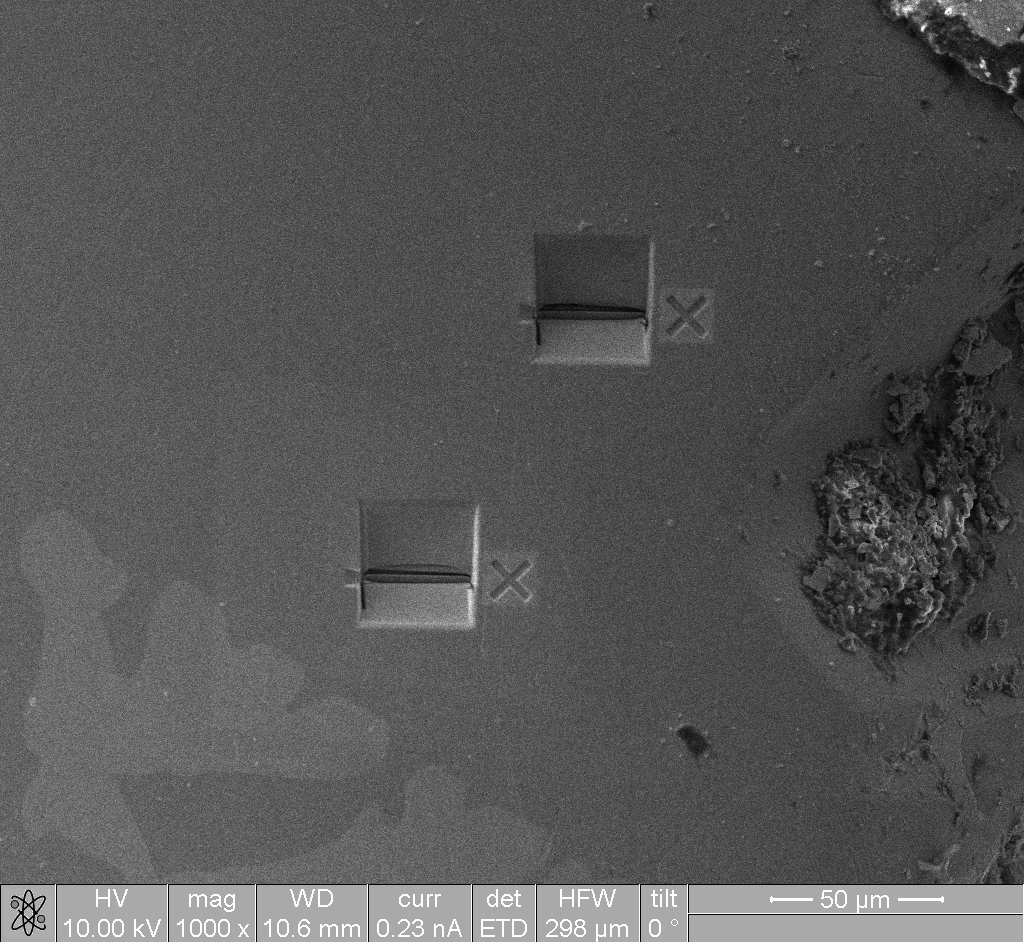}
		\label{fig:bulksample}
	\end{subfigure}%
	\begin{subfigure}[t]{0.47\textwidth}
		\subcaption{}
		\includegraphics[width=0.85\linewidth]{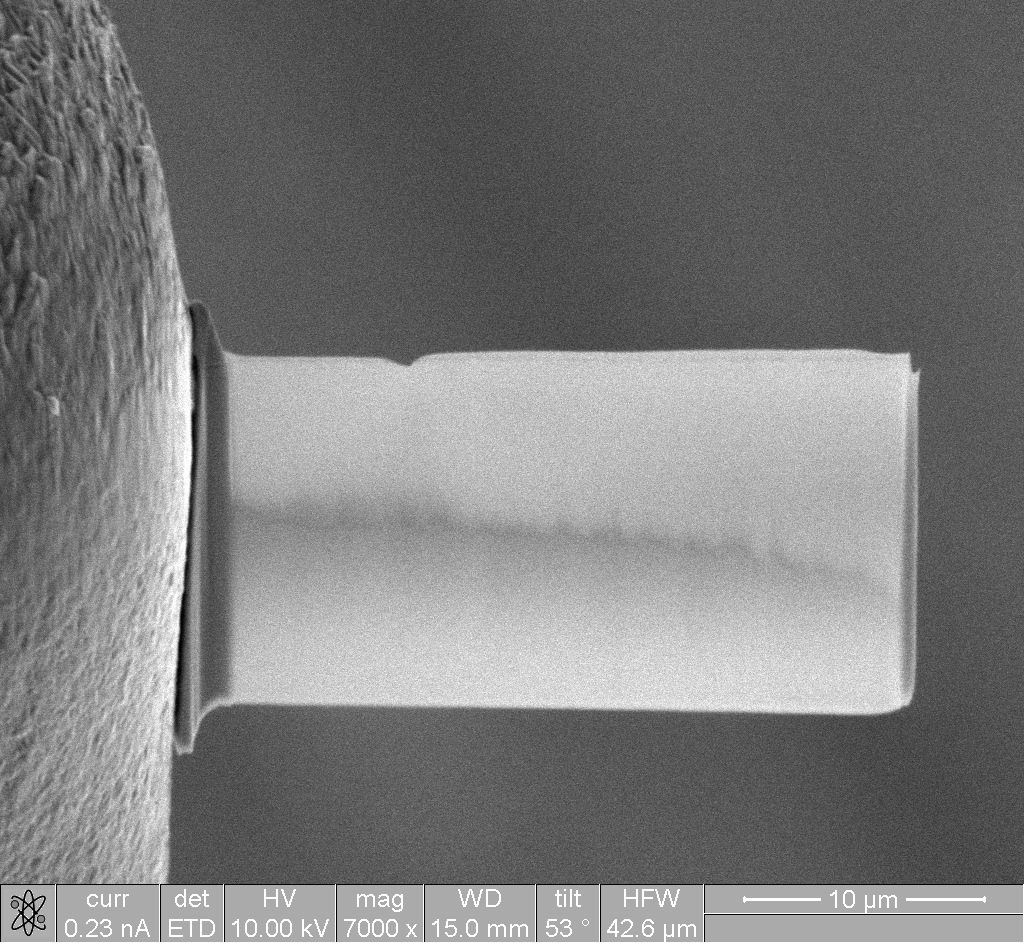}
		\label{fig:lamella}
	\end{subfigure}\\
	\vspace{0.5cm}
	\begin{subfigure}[t]{0.47\textwidth}
		\subcaption{}
		\includegraphics[width=0.85\linewidth]{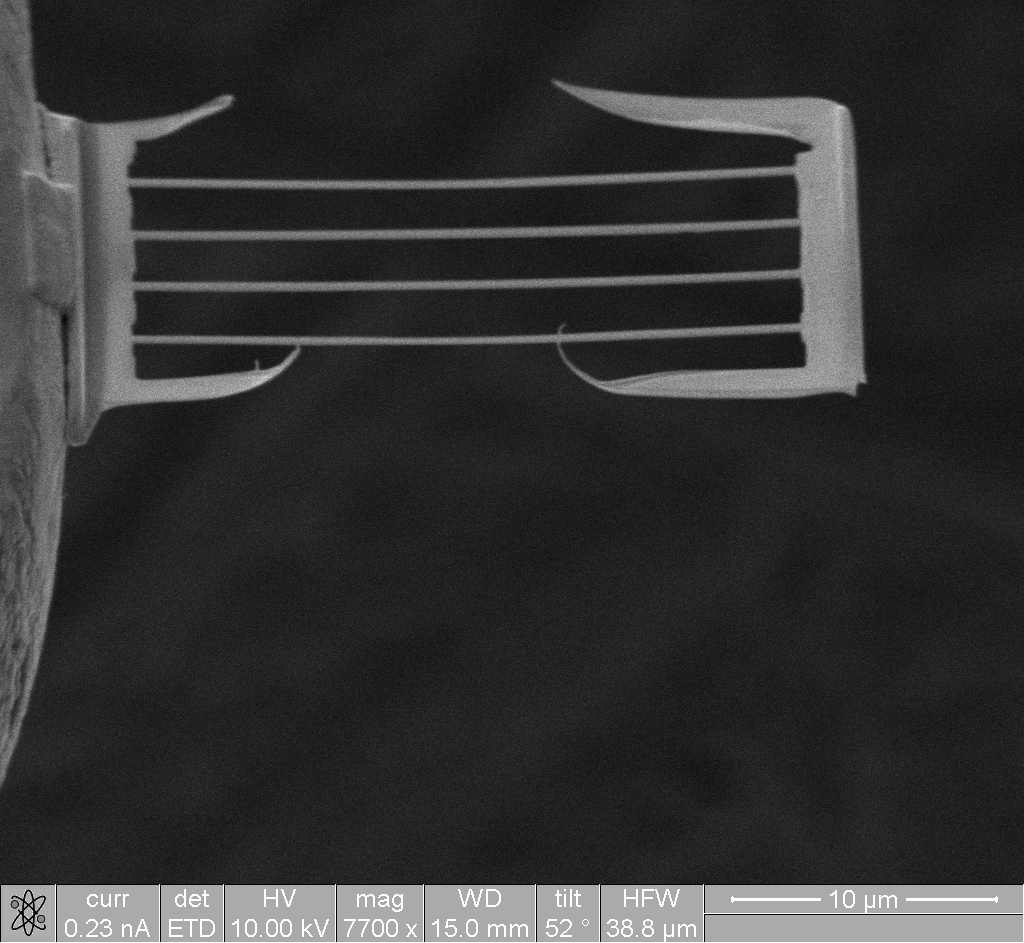}
		\label{fig:mesowires}
	\end{subfigure}%
	\begin{subfigure}[t]{0.47\textwidth}
		\subcaption{}
		\includegraphics[width=0.85\linewidth]{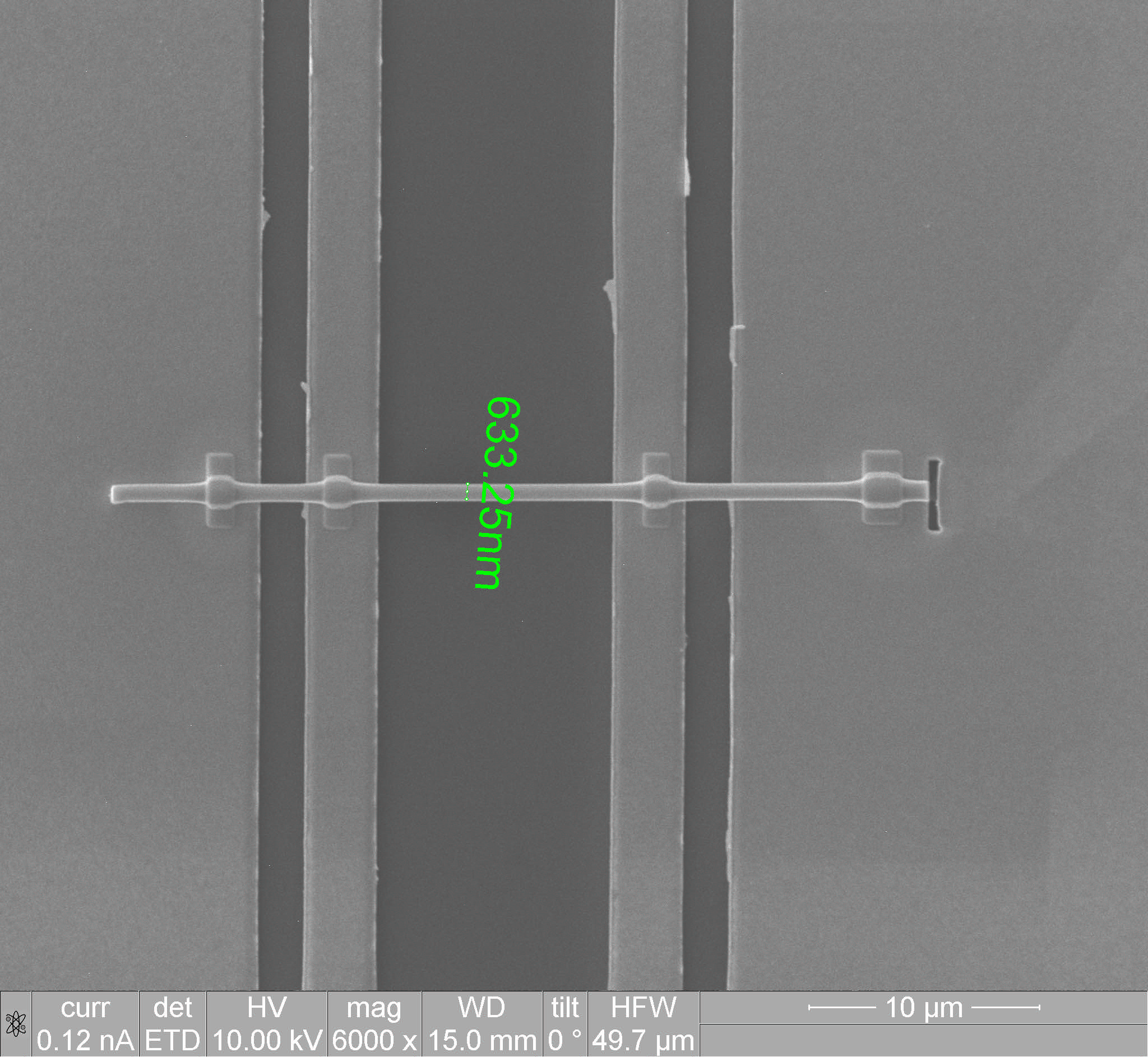}
		\label{fig:NW_contacted}
	\end{subfigure}
\caption{Scanning electron microscopy images of (\textbf{a}) the bulk single crystal with cut out lamellas; (\textbf{b})~a lamella; (\textbf{c}) mesowires cut from the lamella; and (\textbf{d}) a single mesowire on the measuring~platform.}
\label{fig3}
\end{figure}

The mesowires were fabricated with a top-down approach, in which the Ga$^+$ focused ion beam (FIB) of a FEI Quanta 200 3D Dual Beam system is used to form mesowires out of bulk single crystals. To protect the sample surface, first a platinum metal layer is deposited over the area of interest. Two trenches, one on each side of the metal line, are then cut with the FIB, so that a thin platelet, a so-called lamella, is obtained. In order to remove the lamella from the surrounding bulk crystal, it is tilted and cut free except for a small area on the side. It is then welded to a micro-manipulator needle using ion beam induced tungsten deposition. Finally, the remaining bridge between the lamella and the bulk sample is cut, and the lamella is transferred to and mounted on a transmission electron microscope (TEM) grid (Figure~\ref{fig3}b), where the FIB is used to shape thin wires out of it (Figure~\ref{fig3}c). The amorphous surface layer that forms due to bombardment with the high-energy Ga$^+$ ions is removed by an Ar$^+$ soft-milling procedure once the mesowires have almost the desired thickness. Next, the individual wires are again welded to the micro-manipulator tip and cut off from the rest of the material. The samples are then transported to a measuring platform, which was prepared beforehand using optical lithography. It consists of Ti/Au stripes (5~nm/45~nm thick, respectively) on an etched Si/SiO$_2$ substrate and can hold up to four wires. The wires are contacted there, using a combination of ion beam induced and electron beam induced tungsten deposition (Figure~\ref{fig3}d).

\subsection{Measurement Method}
\begin{figure}[H]
	\begin{subfigure}[t]{0.4\textwidth}
         \subcaption{}
		 \includegraphics[height=5cm]{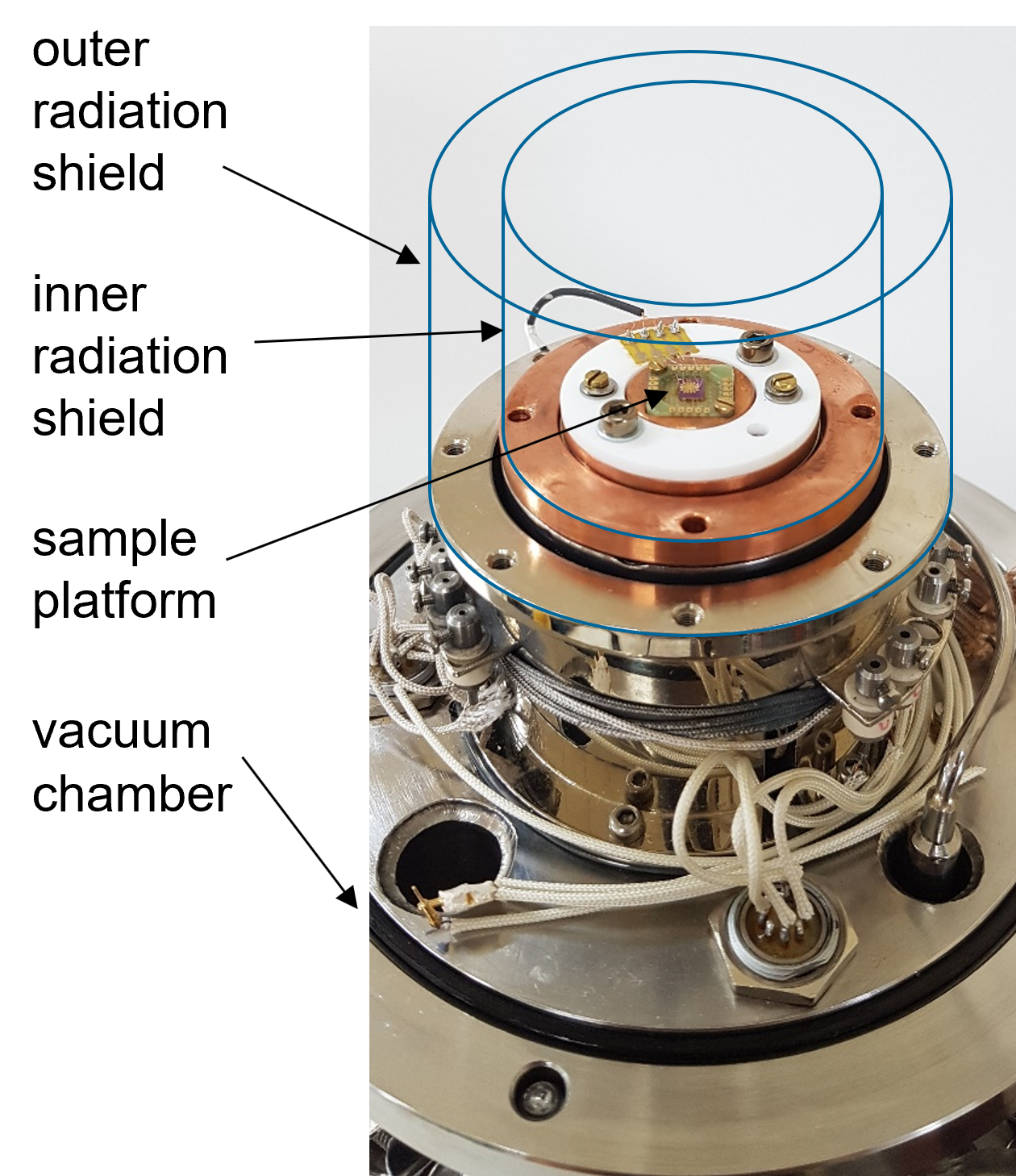}
		\label{fig:setup}
	\end{subfigure}%
	\begin{subfigure}[t]{0.5\textwidth}
		\subcaption{}
		 \includegraphics[height=5cm,trim={0cm 0cm 0cm 0cm},clip]{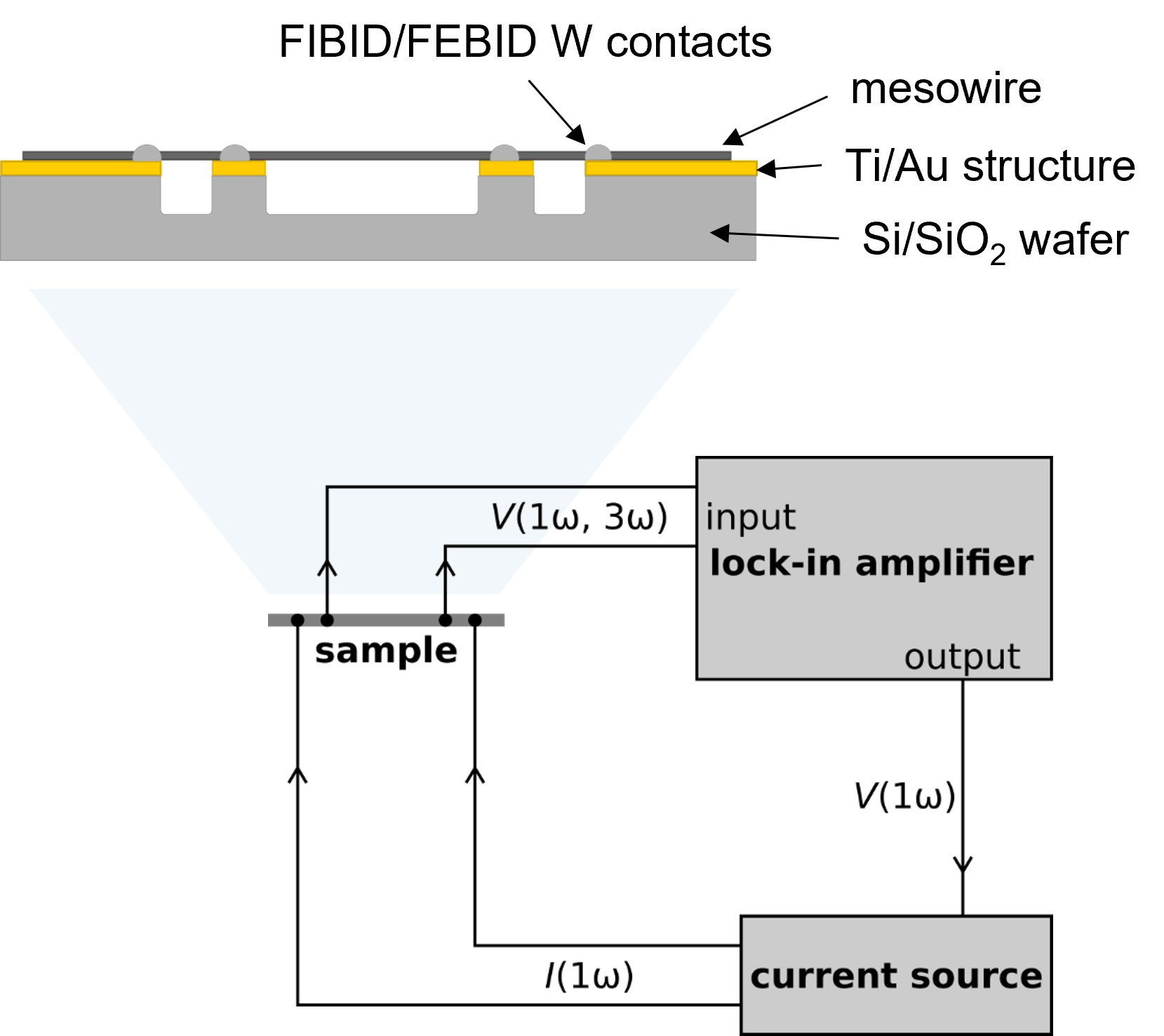}
	    \label{fig:circuit}
	\end{subfigure}\\
	\caption{(\textbf{a}) measurement setup; (\textbf{b}) schematic of the measurement circuit, adapted from \cite{Lu2001}.}
	\label{fig4}
\end{figure}

The mesowires were measured at temperatures between 300~K and 80~K in an LN$_2$-flow cryostat, which is equipped with two radiation shields inside a vacuum chamber. Using a 4-point configuration, a $3\omega$ method was deployed to measure the sample's thermal conductivity. In this technique, which is commonly used for microwires \cite{Bhatta2010}, nanotubes~\cite{Choi2005} and nanowires \cite{Chien2016,Li2014}, the suspended sample acts as both heater and thermometer. This makes it less sensitive to radiation losses than other measurement methods, which is a big advantage for high-temperature measurements. Figure~\ref{fig4}b shows a schematic of the set-up: a lock-in amplifier (Signal Recovery 7265) provides a sinusoidal voltage signal, which is turned into an AC current using a current source (Anmesys AMS220). The current is fed through the sample, and the voltage is picked up on the inner contacts. The $V(I)$ dependence of the first harmonic signal provides information on the electrical resistance, whereas the thermal conductivity can be obtained from the third harmonic using the equation \cite{Lu2001}

\begin{equation}
	V_{3\omega} \approx \frac{4I^3 RR'L}{\pi^4 A\kappa\sqrt{1+(2\omega\gamma)^2}}\, ,
	\label{eq:3omega}
\end{equation}

\noindent 
where $I$ is the root mean square of the applied AC current, $\omega$ is its angular frequency, $R$ and $R'$ are the sample's resistance and its temperature derivative, $\gamma$ is the thermal time constant and $L$ and $A$ are the length between the voltage contacts and the cross-section of the sample, respectively. This equation not only gives direct information on the thermal conductivity $\kappa$, but also via the thermal time constant $\gamma$, on the specific heat $C_{\rm P}$.

\begin{figure}[H]
    \captionsetup[subfigure]{font={bf,small}, skip=1pt, margin= -1pt, singlelinecheck=false}
	\begin{subfigure}[t]{0.55\textwidth}
		  \subcaption{}
		 \includegraphics[height=5cm,trim={0cm 0cm 0cm 0cm},clip]{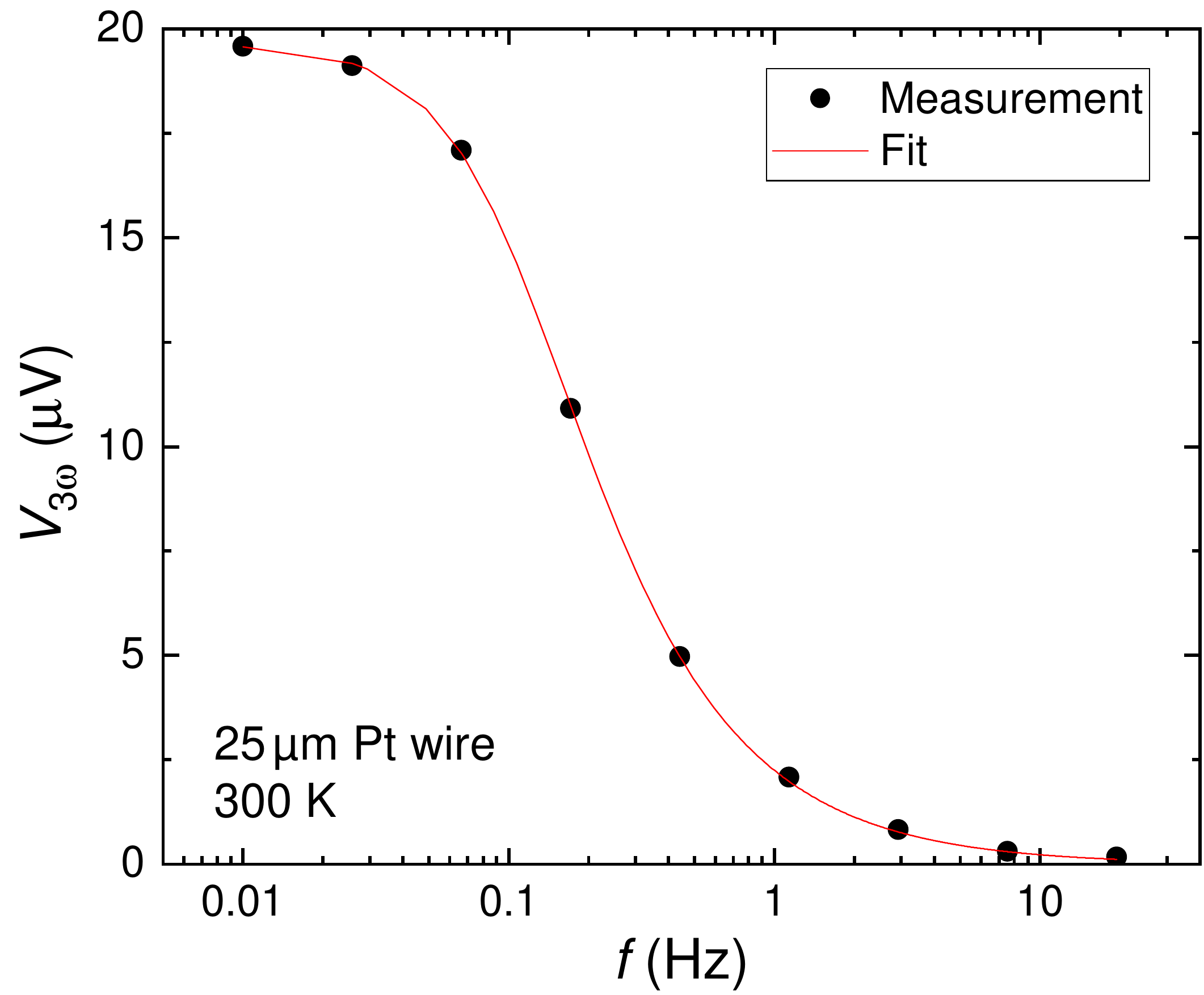}
		\label{fig:Pt}
	\end{subfigure}%
	\begin{subfigure}[t]{0.55\textwidth}
		\subcaption{}
		 \includegraphics[height=5cm,trim={0cm 0cm 0cm 0cm},clip]{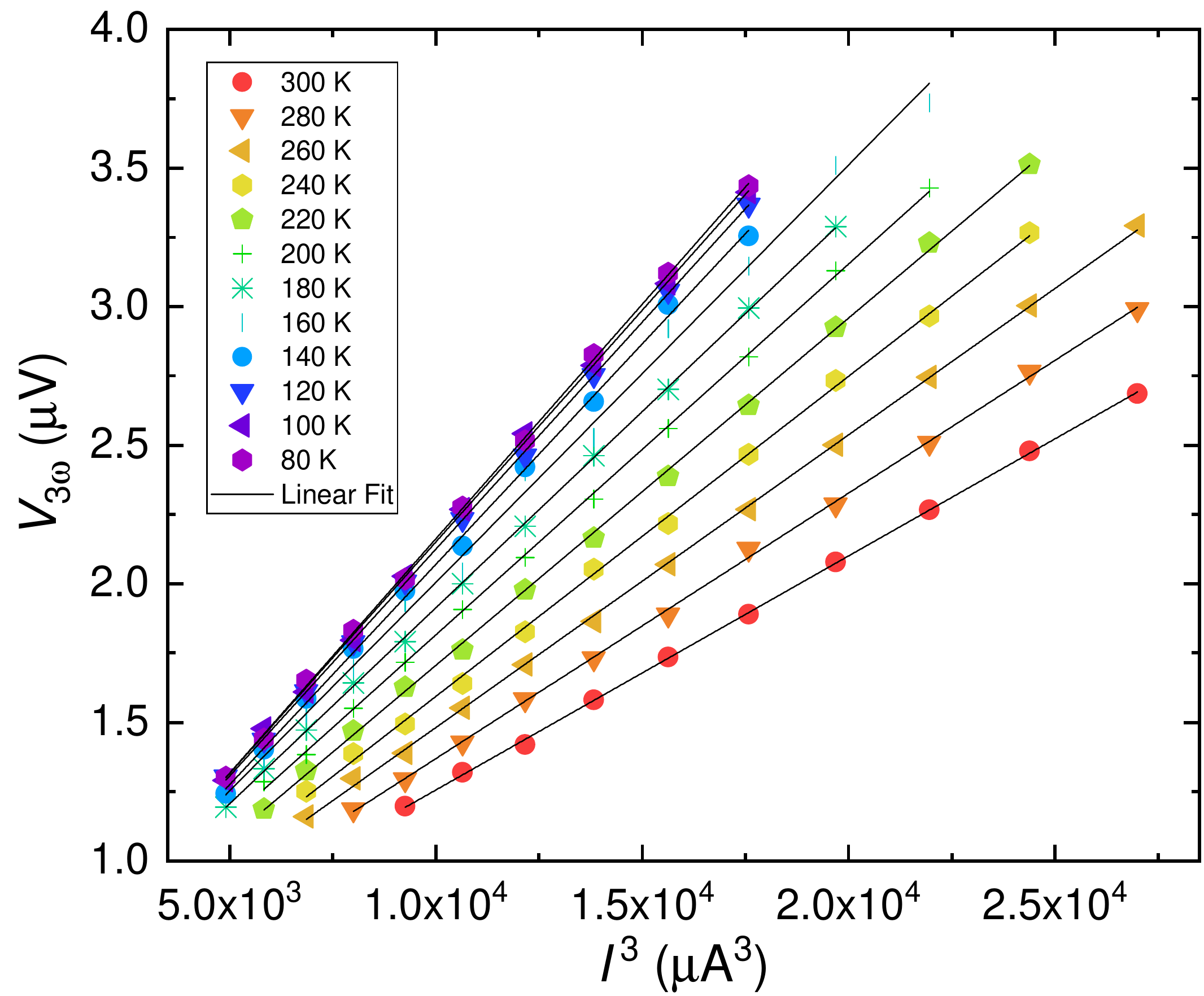}
		\label{fig:V3w_vs_I3}
	\end{subfigure}
	\caption{(\textbf{a}) Frequency dependence of the third harmonic voltage signal of a Pt test wire at 300~K, measured with 3.5\,mA; (\textbf{b}) current dependence of the third harmonic of a 920~nm La-BAS mesowire at different temperatures. Prior to measuring the $V(I)$ dependencies, frequency sweeps were performed to select a measuring frequency at which the $3\omega$ signal is nearly frequency independent and thus Equation (\ref{eq:3omega_low-f}) is valid. For better visibility, not all measured temperatures are included in this plot.}
	\label{fig5}
\end{figure}

\textls[+10]{The measurement setup and program were tested using a 25\,$\upmu$m thick Pt wire. The frequency dependent third harmonic at 300~K is shown in Figure \ref{fig5}a. The fitted curve yields $\kappa = 71.9~$\,Wm$^{-1}$K$^{-1}$, which is in excellent agreement with the literature value for Pt, \mbox{$\kappa=71.6$}\,~Wm$^{-1}$K$^{-1}$ \cite{CRC2005}. The difference between the obtained specific heat (\mbox{$C_{\rm P}=137~$\,Jkg$^{-1}$K$^{-1}$})} and the literature value ($C_{\rm P}=133$\,Jkg$^{-1}$K$^{-1}$ \cite{CRC2005}) is slightly larger, but still only about 3\%. With these results, we deemed the setup suitable for the mesowire measurements.

Compared to the microwire measurements, the characteristic frequency dependence of the mesowires is shifted to higher frequencies. As it can no longer be fully detected by the lock-in amplifier, we decided to use a different approach for determining the thermal conductivity: in the low-frequency limit, Equation (\ref{eq:3omega}) takes the form

\begin{equation}
	V_{3\omega} \approx \frac{4I^3 RR'L}{\pi^4 A\kappa}\, ,
	\label{eq:3omega_low-f}
\end{equation}

\noindent
i.e., the third harmonic becomes nearly frequency independent. Although the information on the specific heat is lost, the thermal conductivity can be readily obtained by varying the excitation current at a fixed frequency. Figure \ref{fig5}b shows that the third harmonic signal of a 920~nm mesowire is linear in $I^3$ for the whole temperature range. The slopes of the linear fits are used to calculate $\kappa$.

\section{Results}

\begin{figure}[H]
    \captionsetup[subfigure]{font={bf,small}, skip=1pt, margin= -1pt, singlelinecheck=false}
	\begin{subfigure}[t]{0.55\textwidth}
		  \subcaption{}
		 \includegraphics[height=5cm,trim={0cm 0cm 0cm 0cm},clip]{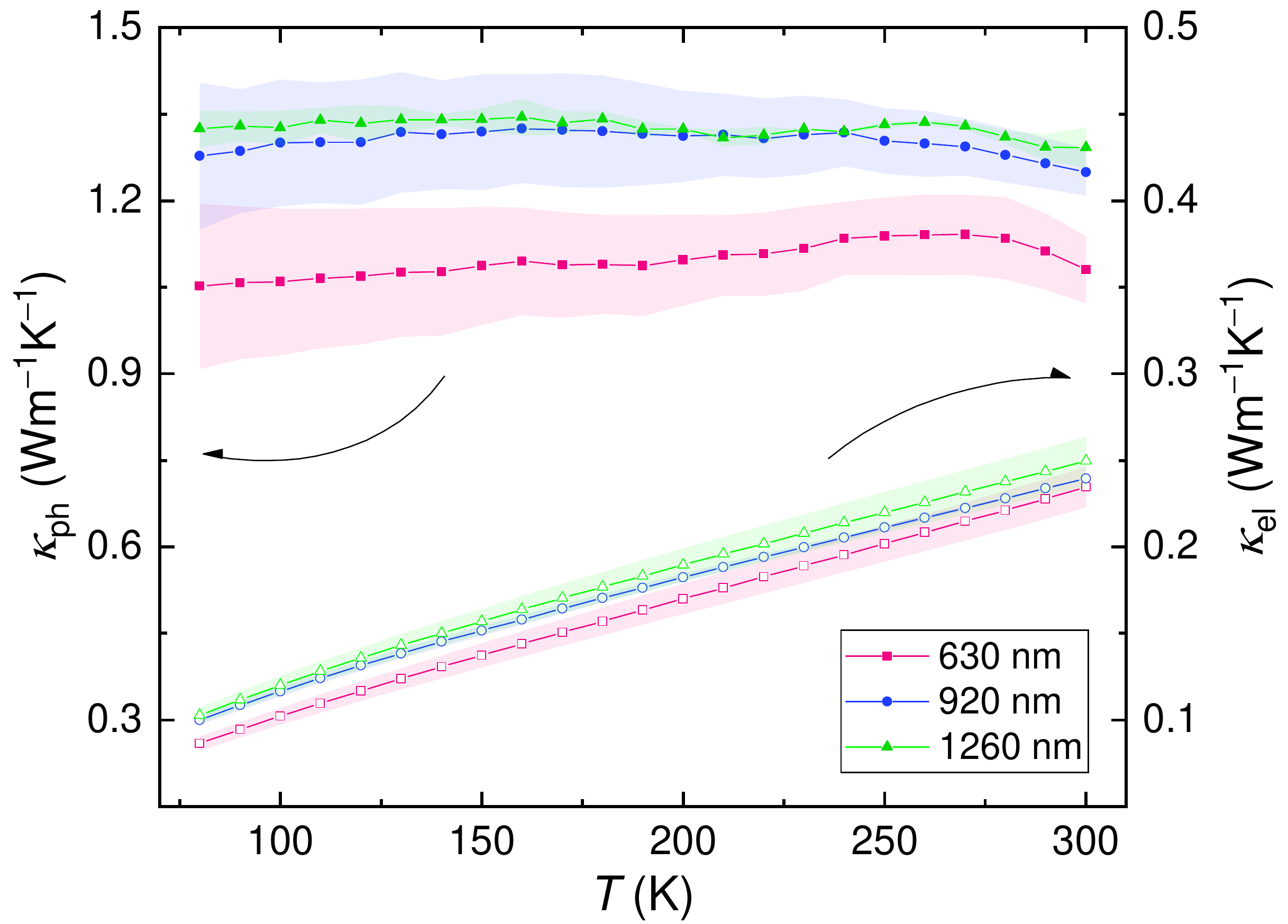}
		\label{fig:kappaT}
	\end{subfigure}%
	\begin{subfigure}[t]{0.55\textwidth}
		\subcaption{}
		 \includegraphics[height=5cm,trim={0cm 0cm 0cm 0cm},clip]{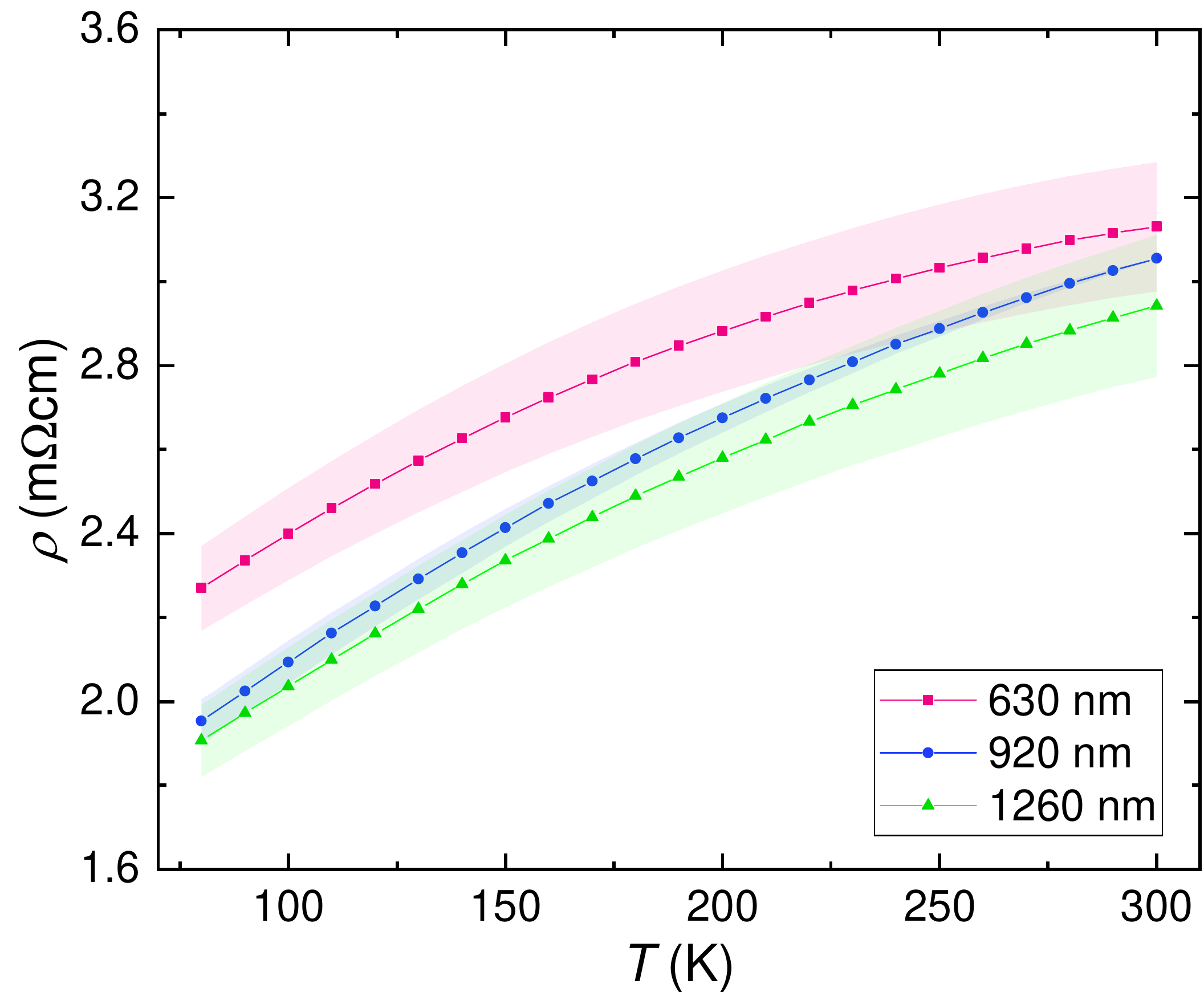}
		\label{fig:rhoT}
	\end{subfigure}
	\caption{Temperature-dependent (\textbf{a}) phonon thermal conductivity (full symbols, left axis) and electron thermal conductivity (open symbols, right axis), as estimated from the Wiedemann--Franz law and (\textbf{b}) electrical resistivity.}
	\label{fig6}
\end{figure}

Figure~\ref{fig6}a shows the temperature-dependent thermal conductivity of La-BAS mesowires with three diameters of roughly 630~nm, 920~nm and 1260~nm. The results are the average of measurements of two different mesowires, the dispersion being represented by the shaded area. The left axis shows the phonon thermal conductivity, which was obtained by subtracting the electron thermal conductivity from the measured total thermal conductivity. It is rather flat over the measured temperature range, with a small decrease at higher temperatures. The room temperature phonon thermal conductivity $\kappa_{\rm ph}$ ranges from $1.29$\,Wm$^{-1}$K$^{-1}$ for the thickest wires to $1.08$\,Wm$^{-1}$K$^{-1}$ for the wires with 630~nm diameter. Compared to the bulk phonon thermal conductivity \cite{Prokofiev2013}, this corresponds to a reduction of roughly 40\% for the thinnest wires.

\textls[-25]{The electron thermal conductivity is plotted on the right axis of Figure~\ref{fig6}a. It was estimated using the Wiedemann--Franz law $\kappa_{\rm el}=LT/\rho$, with the Lorenz number \mbox{$L=2.44 \times 10^{-8}$\,~V$^2$K$^{-2}$}} and $\rho$ as shown in Figure~\ref{fig6}b. With the highest obtained value being $\kappa_{\rm el}\sim 0.25$\,~Wm$^{-1}$K$^{-1}$ at room temperature, it is apparent that the thermal conductivity is governed by the phonon contribution.

As expected for a metal, the electrical resistivity of the La-BAS mesowires decreases with decreasing temperature.
Comparing Figure~\ref{fig6}a,b, one can see that, while the thermal conductivity decreases when reducing the diameter, the electrical resistivity somewhat increases. In terms of the overall thermoelectric conversion efficiency, the question of whether the positive effects outweigh the negative ones will thus have to be addressed.

\section{Discussion}

The phonon thermal conductivity of clathrates can be fitted using a modified Callaway model, as proposed in \cite{Ikeda2019}. It uses the formalism of the well established Callaway model~\cite{Callaway1959}, but accounts for the suppression of high-frequency phonons by substituting the Debye temperature with the Einstein temperature of the lowest-lying rattling mode,

\begin{equation}
    \kappa_{\rm ph}=\dfrac{k^4_\mathrm{B}T^3}{2\pi^2v_{\rm s}\hbar^3}\int^{\theta_{\rm E}/T}_0\tau\dfrac{x^4e^x}{(e^x-1)^2} dx \quad \mbox{with} \quad x\equiv\dfrac{\hbar\omega}{k_\mathrm{B}T}\, ,
    \label{eq:mCallaway}
\end{equation}

\noindent
where $\tau$ is the relaxation rate obtained from three main contributions and Matthiessen's~rule

\begin{equation}
    \begin{aligned}
    \tau^{-1}&=\tau^{-1}_{\rm D}+\tau^{-1}_{\rm U}+\tau^{-1}_{\rm B} \quad \mbox{with}\\
    \tau^{-1}_{\rm D}&=A\omega^4\, , \quad
    \tau^{-1}_{\rm U}=B\omega^2T\, , \quad
    \tau^{-1}_{\rm B}=v_{\rm s}/\lambda\, .
    \label{eq:Matthiessen}
    \end{aligned}
\end{equation}

The contributions $\tau^{-1}_{\rm D}, \tau^{-1}_{\rm U}$ and $\tau^{-1}_{\rm B}$ are the defect scattering rate, the high-temperature Umklapp scattering rate according to Klemens \cite{Klemens1958} and the boundary scattering rate, respectively.

\begin{figure}[H]
    \captionsetup[subfigure]{font={bf,small}, skip=1pt, margin= -1pt, singlelinecheck=false}
	\begin{subfigure}[t]{0.55\textwidth}
      \subcaption{}
		 \includegraphics[height=5cm,trim={0cm 0cm 0cm 0cm},clip]{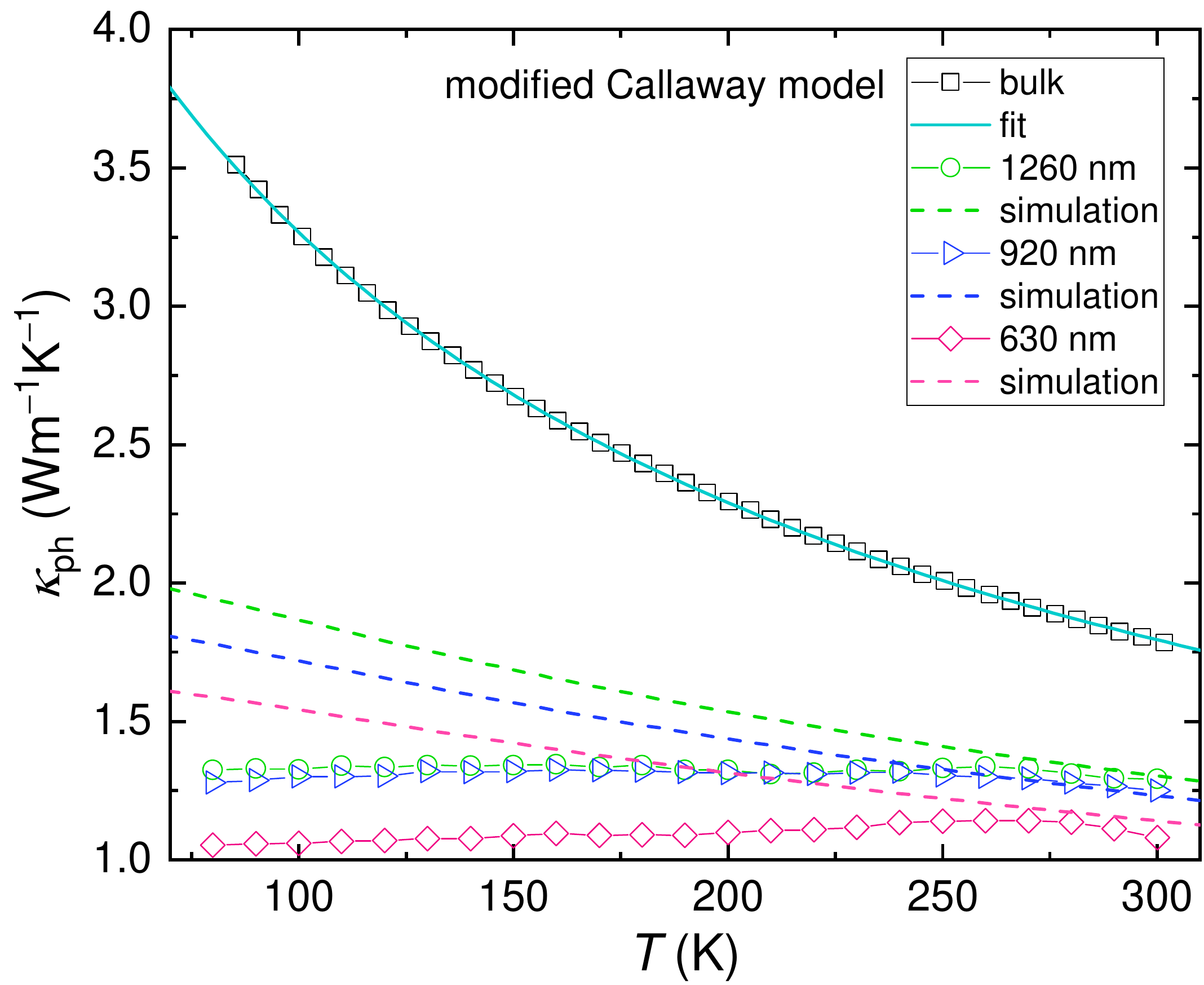}
		\label{fig:mCallaway}
	\end{subfigure}%
	\begin{subfigure}[t]{0.55\textwidth}
		 \subcaption{}
		 \includegraphics[height=5cm,trim={0cm 0cm 0cm 0cm},clip]{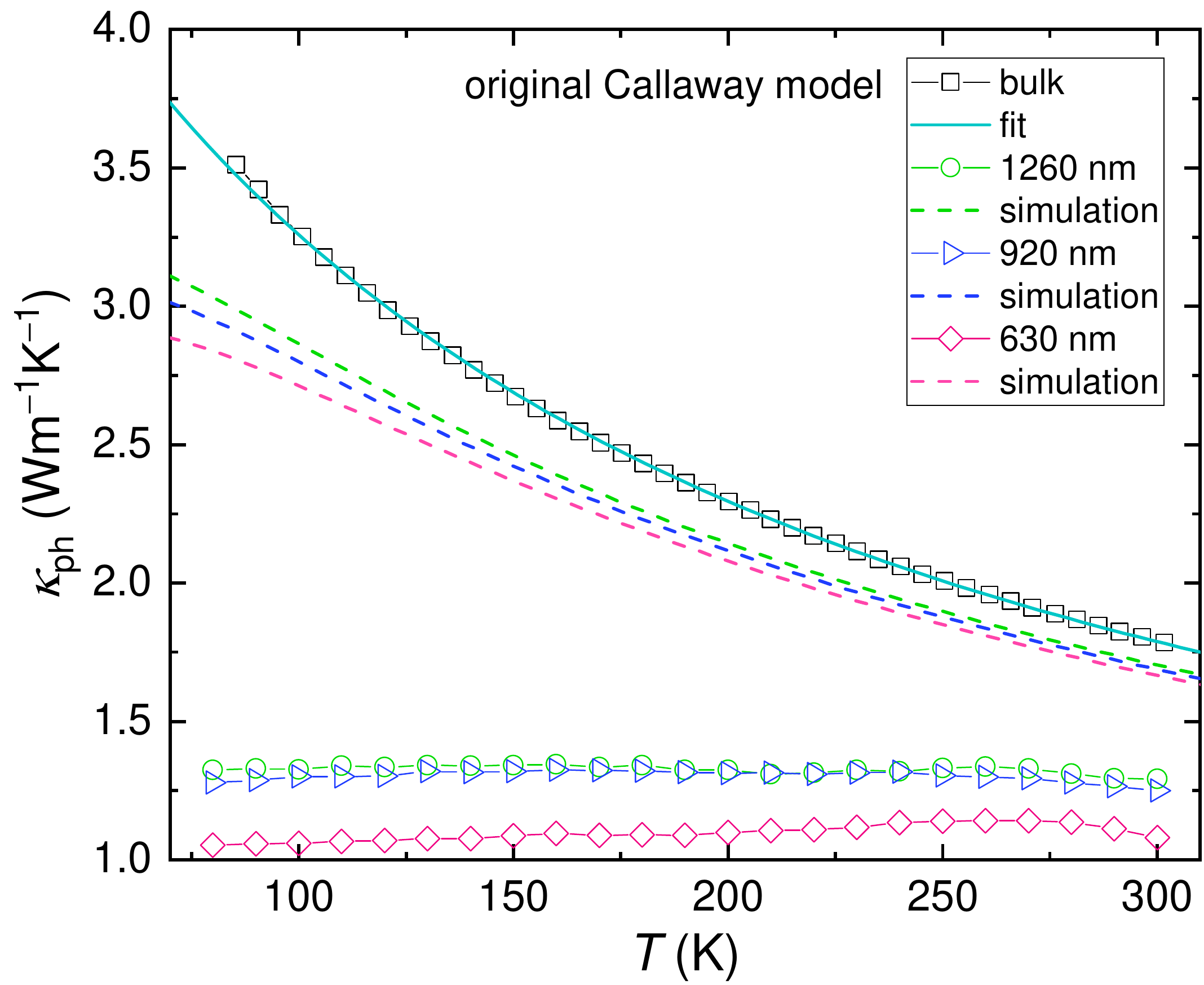}
	\label{fig:oCallaway}
	\end{subfigure}\\
	\caption{Temperature-dependent phonon thermal conductivity of a bulk La-BAS single crystal and mesowires fabricated from the same batch (symbols), the best fit to the bulk single crystal (full blue line) and simulated curves (dashed lines) for which the boundary scattering length is fixed to the mesowire diameters and all other parameters are kept unchanged, using (\textbf{a}) the modified Callaway model with $\theta_{\rm E}=82.3$~K and (\textbf{b}) the original Callaway model with $\theta_{\rm D}=348.3$~K. Data for the bulk single crystal are taken from \cite{Ikeda2015}. Only the modified Callaway model can capture the overall high-temperature behavior. Deviations at low temperatures are attributed to small changes in the defect scattering rate due to the FIB processing.}
	\label{fig7}
\end{figure}

Using Equations (\ref{eq:mCallaway}) and (\ref{eq:Matthiessen}), and the values $\theta_{\rm E}=82.3$~K and $v_{\rm s}=3330$\,m/s \cite{Ikeda2015}, a good fit to the bulk phonon thermal conductivity of La-BAS is obtained, as shown in Figure~\ref{fig7}a. From the fit, one obtains the defect scattering constant $A=5.74\times 10^{3}$~K$^{-4}$s$^{-1}$, the Umklapp scattering constant $B=38.8\times 10^{3}$~K$^{-3}$s$^{-1}$ and the boundary scattering length $\lambda=4.11\times 10^{-5}$\,m.

We then simulate curves assuming that the FIB processing changed the boundary scattering length to the diameters of the mesowires (1260~nm, 920~nm and 630~nm) but left the other scattering processes unaffected. These curves match the measured data quite well at high temperatures. Deviations are observed below about 270~K. They are likely due to additional defects introduced during the mesostructuring process. As shown in \cite{Ikeda2019}, defect scattering in type-I clathrates dominates at low temperatures but can be discarded as the dominating effect at room temperature.

For comparison, we also fit the bulk data with the original Callaway model with $\theta_{\rm D}=348.3$~K \cite{Ikeda2015} (Figure~\ref{fig7}b). This fit, though it naturally yields different fitting parameters, is of similar quality as the one using the modified Callaway model. However, the simulated curves, obtained by again changing the boundary scattering length to match the mesowire diameters, differ much less from the bulk curve and fail to capture the high-temperature mesowire data. This confirms the validity of the assumption that the (Kondo-like) interaction between the acoustic modes and the rattling modes effectively cuts off the acoustic phonon spectrum and thus shifts the weight of heat-carrying phonons to lower wavelength~ones.

\begin{figure}[H]
	 \includegraphics[width=10cm,trim={0cm 0cm 0cm 0cm},clip]{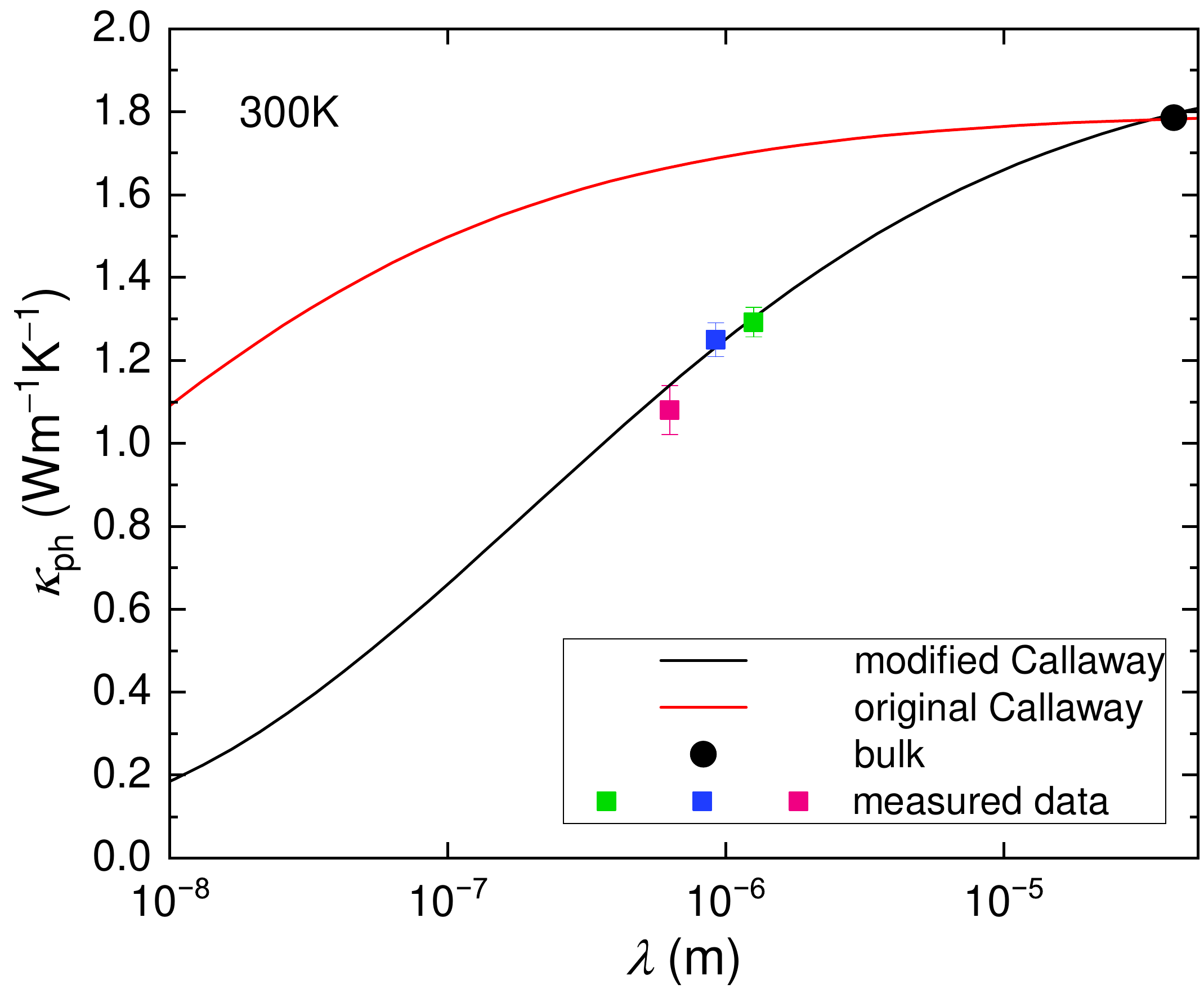}
	\caption{Phonon thermal conductivity as a function of boundary scattering length at 300~K, simulated using the modified (black) and the original Callaway model (red). Squares denote mesowire data, the circle the bulk data point taken from \cite{Ikeda2015}.}
	\label{fig:CallawayL}
\end{figure}

Focusing on the high-temperature region, where effects of point defect scattering were shown to be unimportant \cite{Ikeda2019}, we visualize the difference between these two models by plotting the phonon thermal conductivity vs. the boundary scattering length. Figure~\ref{fig:CallawayL} shows the respective simulations at 300~K, together with the data points for bulk and mesowires. It is clear that only the modified Callaway model, where the Einstein temperature takes the role of the Debye temperature, captures our data. The fact that the boundary-scattering effect is governed by the Einstein and not by the Debye temperature makes type-I clathrates uniquely suited for size effect approaches. Moving on, it will be of great interest to test whether the thermal conductivity of clathrate mesowires with a lower Einstein temperature does in fact show an even stronger size dependence.

\begin{figure}[H]
	 \includegraphics[width=10cm,trim={0cm 0cm 0cm 0cm},clip]{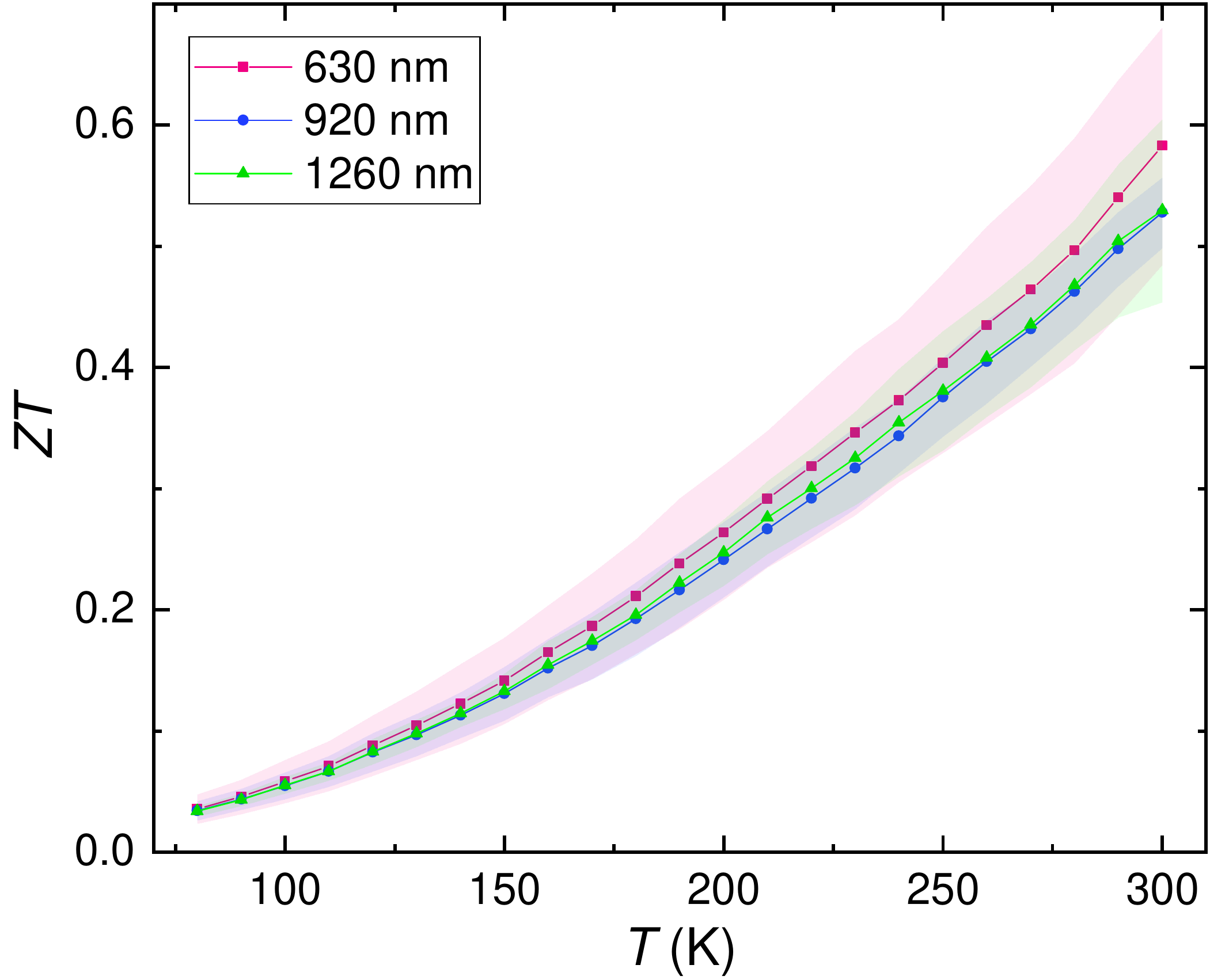}
	\caption{$ZT$ as a function of $T$. The behavior of the mesowires with a 630~nm diameter at high temperatures points towards an improvement as a result of mesostructuring.}
	\label{fig:ZT}
\end{figure}

Finally, we estimate the effect of the mesostructure on the thermoelectric efficiency of La-BAS. Using the compound's Seebeck coefficient measured in \cite{Prokofiev2013}, we calculate the figure of merit, as shown in Figure~\ref{fig:ZT}. While $ZT$ is almost identical for the mesowires with diameters of 1260~nm and 920~nm, a notable enhancement is observed for the thinnest mesowires (of 630~nm diameter), caused by the strong reduction of phonon thermal conductivity at this diameter.

For materials with more conventional phonon spectra, where the Debye temperature governs the phonon thermal conductivity, size effects are observed for much shorter dimensions. For instance, in Bi$_2$Te$_3$ wires, a reduction of the room temperature phonon thermal conductivity of only $\sim$$25$\% is observed for a diameter of 300~nm \cite{Munoz2017}, and, in Bi$_2$Se$_3$, a reduction of $\sim$$29$\% was found for a diameter of 200~nm \cite{Dedi2021}. The room temperature total thermal conductivities of 182~nm PbTe wires \cite{Roh2010} and 50--100~nm PbSe wires \cite{Liang2009} show a reduction of $\sim$$50$\%, but, in the latter work, the decrease is partly attributed to the rough surface. While the total thermal conductivity of InAs nanowires of 40~nm diameter shows an 80\% decrease, the decrease expected for 400~nm wires amounts to only $\sim$$26$\%~\cite{Swinkels2015}. Most studies on Si have focused on extremely thin wires \cite{Bou08.1}. In a study comparing somewhat larger structures \cite{Ju2005}, the change in the total thermal conductivity between films with characteristic dimensions of 250~nm and 150~nm amounts to less than 10\%, whereas, for nanowires of 50~nm and 20~nm, it differs by more than 45\%, showing that substantial size effects in Si require nanostructures as opposed to mesostructures. As our work shows, this is in contrast to type-I clathrates, where sizable effects can already be achieved with mesostructures. Our proof-of-concept study opens the door for new approaches to enhance $ZT$ in type-I clathrates, using mesostructures as opposed to nanostructures that are much more difficult to stabilize.

\vspace{6pt} 


\authorcontributions{Investigation, formal analysis and visualization, M.L.; investigation and software, G.L.;  resources, M.T. (measuring platforms), A.S.-T. (mesowires), A.P. (bulk sample); writing---original draft preparation, M.L.; writing---review and editing, G.L. and S.P.; conceptualization, supervision, project administration and funding acquisition, S.P. All authors have read and agreed to the published version of the manuscript.}

\funding{This research was funded by the Austrian Science Fund (FWF), Grant Nos. 29279, W1243 and I5868--FOR5249.}

\dataavailability{Original data are available from the corresponding authors upon reasonable request.} 

\acknowledgments{Open Access Funding by the Austrian Science Fund (FWF).}

\conflictsofinterest{The authors declare no conflict of interest.} 






\begin{adjustwidth}{-\extralength}{0cm}

\reftitle{References}

\PublishersNote{}
\end{adjustwidth}
\end{document}